\documentclass[aps,prb,twocolumn,showpacs,amsmath,amssymb]{revtex4}
\usepackage{dcolumn}
\usepackage{bm}
\usepackage{graphicx}
\usepackage{subfigure}
\usepackage{times}
\usepackage{epstopdf}
\usepackage[makeroom]{cancel}
\usepackage[usenames,dvipsnames,svgnames]{xcolor}

\begin{document}
\title{Superconducting order parameter $\pi$-phase shift in magnetic impurity wires}
\author{Kristofer Bj\"{o}rnson$^1$, Alexander. V. Balatsky$^{2,3}$, and Annica M. Black-Schaffer$^1$}
\affiliation{$^1$Department of Physics and Astronomy, Uppsala University, Box 516, S-751 20 Uppsala, Sweden}
\affiliation{$^2$Nordita, Center for Quantum Materials, KTH Royal Institute of Technology, and Stockholm University, Roslagstullsbacken 23, S-106 91 Stockholm, Sweden}
\affiliation{$^3$Institute for Materials Science, Los Alamos National Laboratory, Los Alamos, NM 87545, USA}

\begin{abstract}
It has previously been found that a magnetic impurity in a conventional $s$-wave superconductor can give rise to a local $\pi$-phase shift of the superconducting order parameter.
By studying a finite wire of ferromagnetic impurities, we are able to trace the origin of the $\pi$-phase shift to a resonance condition for the  Bogoliubov-de Gennes quasiparticle states. When non-resonating states localized at the impurity sites are pulled into the condensate for increasing magnetic strength, the superconducting order parameter is reduced in discrete steps, eventually resulting in a $\pi$-phase shift. We also show that for a finite spin-orbit coupling, the $\pi$-phase shift is preserved and occurs in a large portion of the topologically non-trivial phase. 
\end{abstract}

\maketitle

\section{Introduction}
The superconducting state differs in one major regard from the normal state of a metal in that it without resistance can carry current, even in the presence of imperfections in the crystal.
In fact, conventional fully gapped $s$-wave superconductors are remarkably stable and unaffected by potential and other general time-reversal-invariant disorder, as established by the Anderson theorem.\cite{JPhysChemSolids.11.26}
However, for magnetic impurities, large enough concentrations are detrimental for superconductivity and even a single magnetic impurity notably modifies the properties of the superconductor locally around the impurity.
The study of individual impurities can even reveal important features of the superconducting condensate itself.\cite{RevModPhys.78.373}

It is well-known that magnetic impurities in conventional $s$-wave superconductors give rise to localized intra-gap states, called  Yu-Shiba-Rusinov (YSR) states.\cite{ActaPhysSin.21.75, ProgTheorPhys.40.435, JETPLett.9.85}
With increasing magnetic impurity strength, two YSR states leave the energy gap edges and progress into the energy gap as quasiparticle excitations of the superconducting condensate. 
At some critical impurity strength, the YSR states cross each other at the Fermi level and in the process induce a quantum phase transition.
The new ground state is a paired condensate for all but the impurity bound state, which results in the magnetic moment being reduced by $\frac{1}{2}$. 
Previous numerical calculations have also revealed that this YSR state crossing can be associated with an intriguing $\pi$-shift in the phase of the local superconducting order parameter.\cite{PhysRevB.55.12648, PhysRevLett.78.3761, RevModPhys.78.373, PhysRevB.92.064503}
However, so far the origin of this phenomenon has not been fully understood.

In this work we study a finite one-dimensional line of magnetic impurities, on which the individual YSR states can be understood to hybridize and form YSR bands.\cite{PhysRevB.88.155420, PhysRevLett.114.236803, arXiv:1605.00696}
We show that the absolute value of the superconducting order parameter on the wire is reduced in discrete steps every time an individual YSR state crosses the Fermi level. For sufficient many crossings, this results in a change of sign in the superconducting order parameter along the wire, which generalizes the single impurity $\pi$-shift to impurity wire configurations.  
We trace the origin of the $\pi$-shift to a certain type of resonance among the Bogoliubov-de Gennes quasiparticles; it is a result of quasiparticle states which are out-of-phase with the condensate being pulled down below the Fermi level, and thus contributing to the superconducting order parameter.
In fact, we show that it is possible to define a corresponding resonance energy for any quasiparticle state, which contributes negatively or positively to the total energy of the ground state depending on whether it is in- or out-of-phase with the condensate, respectively.
Motivated by the recent interest in Rashba spin-orbit coupled $s$-wave superconductors in the context of topological superconductivity,\cite{PhysRevLett.105.077001, PhysRevLett.105.177002, Science.336.1003, NatPhys.8.795, NatPhys.8.887, Science.346.602, arXiv:1505.06078, PhysRevLett.115.197204, PhysRevB.88.020407, PhysRevB.84.195442}
we also investigate the effect of including a Rashba spin-orbit interaction in the superconductor.
We find that the discrete steps become smoother, but that a superconducting order parameter $\pi$-shift is still present in a large portion of the topologically non-trivial phase.
We also carefully study the superconducting phase in the vicinity of the wire, which we find to be almost always completely unaffected and fixed at zero, apart from in a very narrow range around the $\pi$-phase transition in the wire, where we find small phase gradients in the superconducting order parameter. These gradients are independent on spin-orbit coupling and thus cannot be related to previously detected persistent supercurrents around magnetic impurities.\cite{PhysRevLett.115.116602, PhysRevB.92.214501}

\section{Model}
In order to study a generic conventional $s$-wave superconductor with ferromagnetic impurities, we consider the following two-dimensional model Hamiltonian\cite{PhysRevLett.103.020401, PhysRevB.82.134521, PhysRevB.84.180509, PhysRevB.88.024501, PhysRevLett.115.116602, PhysRevLett.104.040502, PhysRevB.91.214514, PhysRevB.92.214501, arXiv:1605.00696}
\begin{align}
	\mathcal{H} &= \mathcal{H}_{kin} + \mathcal{H}_{sc} + \mathcal{H}_{V_z},
\label{Equation:Tight_binding_Hamiltonian} \\ 
	\mathcal{H}_{kin} &= -t\sum_{\langle\mathbf{i},\mathbf{j}\rangle,\sigma}c_{\mathbf{i}\sigma}^{\dagger}c_{\mathbf{j}\sigma} - \mu\sum_{\mathbf{i},\sigma}c_{\mathbf{i}\sigma}^{\dagger}c_{\mathbf{i}\sigma}, \nonumber \\ 
	\mathcal{H}_{so} &= \alpha\sum_{\mathbf{i}\mathbf{b}}\left(e^{i\theta_{\mathbf{b}}}
		c_{\mathbf{i}+\mathbf{b}\downarrow}^{\dagger}c_{\mathbf{i}\uparrow} + {\rm H.c.}\right) \nonumber \\
	\mathcal{H}_{sc} &= \sum_{\mathbf{i}}\left(\Delta(\mathbf{i}) c_{\mathbf{i}\uparrow}^{\dagger}c_{\mathbf{i}\downarrow}^{\dagger} + {\rm H.c.}\right), \nonumber \\
	\mathcal{H}_{V_z} &= -\sum_{\mathbf{i},\sigma,\sigma'}V_z(\mathbf{i})\left(\sigma_z\right)_{\sigma\sigma'}c_{\mathbf{i}\sigma}^{\dagger}c_{\mathbf{i}\sigma'}, \nonumber
\end{align}
where $c_{\mathbf{i}\sigma}^{\dagger}$ ($c_{\mathbf{i}\sigma}$) are creation (annihilation) operators on a two-dimensional square lattice. Here $t$ is the nearest neighbor hopping parameter and $\mu$ the chemical potential, which generates a generic kinetic energy for the superconductor. For concreteness we set $t = 1$ and $\mu = 0$, but qualitatively, our results are not sensitive to the particular choice of $\mu$. 
We also allow for a finite Rashba spin-orbit interaction in the superconductor with strength $\alpha$, $\mathbf{b}$ being a vector pointing along the nearest neighbor bonds, and $\theta_{\mathbf{b}}$ being the polar angle of $\mathbf{b}$. We mainly focus on the case $\alpha =0$, but consider a finite spin-orbit coupling in the last section. The superconducting condensate only includes on-site pairs, resulting in a fully isotropic superconducting order parameter in the translationally invariant bulk, while spatial inhomogeneities can still give rise to a site-dependent $\Delta(\mathbf{i})$.
The ferromagnetic impurities we model as local magnetic impurity spins in the classical limit, in order to recover the well-known YSR states in the single impurity limit. The ferromagnetic impurities effectively influence the superconductor through a Zeeman exchange field term, which we assume to take a fixed value $V_z \in [0, 2.56]$, which covers the whole range of YSR state behaviors.

We solve Eq.~\eqref{Equation:Tight_binding_Hamiltonian} on a lattice of size $19\times 36$, with magnetic impurities in a wire segment of length $18$ in the middle of the lattice. We use a four-by-four (per site) Nambu-basis $(\begin{array}{cccc}c_{\mathbf{i}\uparrow} & c_{\mathbf{i}\downarrow} & c_{\mathbf{i}\uparrow}^{\dagger} & c_{\mathbf{i}\downarrow}^{\dagger}\end{array})$, in which the Bogoliubov-de Gennes Hamiltonian takes the form
\begin{align}
	\mathcal{H}_{BdG} =& \left[\begin{array}{cc}
				\mathbf{H}_{0}(\mathbf{i},\mathbf{j})	& \boldsymbol{\Delta}(\mathbf{i}, \mathbf{j})\\
				\boldsymbol{\Delta}^{\dagger}(\mathbf{i}, \mathbf{j})	& -\mathbf{H}_{0}^{T}(\mathbf{i}, \mathbf{j})
			\end{array}\right],
\end{align}
where $\mathbf{H}_{0}(\mathbf{i}, \mathbf{j})$ is the normal state Hamiltonian,
\begin{align}
	\boldsymbol{\Delta}(\mathbf{i}, \mathbf{j}) =& \delta_{\mathbf{i}\mathbf{j}}\otimes\left[\begin{array}{cc}
				0	& \Delta(\mathbf{i})\\
				-\Delta(\mathbf{i})	& 0
			\end{array}\right],
\end{align}
and with the $\nu$th eigenstate denoted by $[\begin{array}{cccc}u_{\nu\mathbf{i}\uparrow} & u_{\nu\mathbf{i}\downarrow} & v_{\nu\mathbf{i}\uparrow} & v_{\nu\mathbf{i}\downarrow}\end{array}]^{T} = [\begin{array}{cc}\mathbf{u}_{\nu\mathbf{i}} & \mathbf{v}_{\nu\mathbf{i}}\end{array}]^{T}$.
For the conceptual discussion we will also find it useful to refer to the corresponding continuum expression, where we simply replace the $\mathbf{i},\mathbf{j}$ dependence with a dependence on spatial vector $\mathbf{x}$.
In order to fully capture the effects of the magnetic impurities on the superconducting state we calculate the superconducting order parameter self-consistently:
\begin{align}
	\label{Equation:Self_consistency_expression}
	\Delta(\mathbf{i}) =& -\frac{V_{sc}}{2}\left(\langle c_{\mathbf{i}\downarrow}c_{\mathbf{i}\uparrow}\rangle - \langle c_{\mathbf{i}\uparrow}c_{\mathbf{i}\downarrow}\rangle\right)\nonumber \\
	=& -\frac{V_{sc}}{2}\sum_{E_{\nu} < E_F} \left(v_{\nu\mathbf{i}\downarrow}^{*}u_{\nu\mathbf{i}\uparrow} - v_{\nu\mathbf{i}\uparrow}^{*}u_{\nu\mathbf{i}\downarrow}\right),
\end{align}
using a fixed on-site pair potential, which we set to $V_{sc} = 1.6$, resulting in $\Delta \approx 0.2$ in the absence of magnetic impurities.

\section{Resonating quasiparticle states}
Before studying the influence of the magnetic impurities, we begin by noting that the existence of a finite superconducting order parameter can be understood as a result of a certain type of resonance between the Bogoliubov-de Gennes quasiparticles.
While the superconducting order parameter is calculated using the full sum in Eq.~\eqref{Equation:Self_consistency_expression},
we can also consider the individual terms $(v_{\nu\mathbf{i}\downarrow}^{*}u_{\nu\mathbf{i}\uparrow} - v_{\nu\mathbf{i}\uparrow}^{*}u_{\nu\mathbf{i}\downarrow})$ in the sum.
For each occupied quasiparticle state, this quantity contributes to the total order parameter of the condensate.
For a total order parameter to develop, some of these individual terms obviously needs to be non-zero, corresponding to fractional occupation of electron pairs.
But it is also important that the electron pairs on average have the same phase, or are in-phase with each other, otherwise the terms will interfere destructively.
In other words, a sizable fraction of occupied quasiparticle states need to be in resonance with each other in order for superconductivity to develop, where with resonance we simply mean that the relative phase $\arg(v_{\nu\mathbf{i}\downarrow}^{*}u_{\nu\mathbf{i}\uparrow} - v_{\nu\mathbf{i}\uparrow}^{*}u_{\nu\mathbf{i}\downarrow})$ is the same for different eigenstates, i.e.~the same phase between the electron and hole components for different eigenstates.
To be able to compare relative phases between different quasiparticle states, we choose the bulk phase $\Delta/|\Delta|$ of the condensate as reference and define the quantity
\begin{align}
	\delta_{\nu}(\mathbf{i}) \equiv |\delta_{\nu}(\mathbf{i})|e^{i\theta_{\nu}(\mathbf{i})} \equiv -\frac{1}{2}\frac{v_{\nu\mathbf{i}\downarrow}^{*}u_{\nu\mathbf{i}\uparrow} - v_{\nu\mathbf{i}\uparrow}^{*}u_{\nu\mathbf{i}\downarrow}}{\Delta/|\Delta|}.
\end{align}
If $\theta_{\nu}(\mathbf{i}) = 0$, we say that the $\nu$th quasiparticle state is in-phase, or in resonance, with the condensate at site $\mathbf{i}$, while $\theta_{\nu}(\mathbf{i}) = \pm\pi$ corresponds to an eigenstate that is completely out-of-phase, or out of resonance, with the condensate at site $\mathbf{i}$.

\begin{figure}[htb]
\includegraphics[width=245pt]{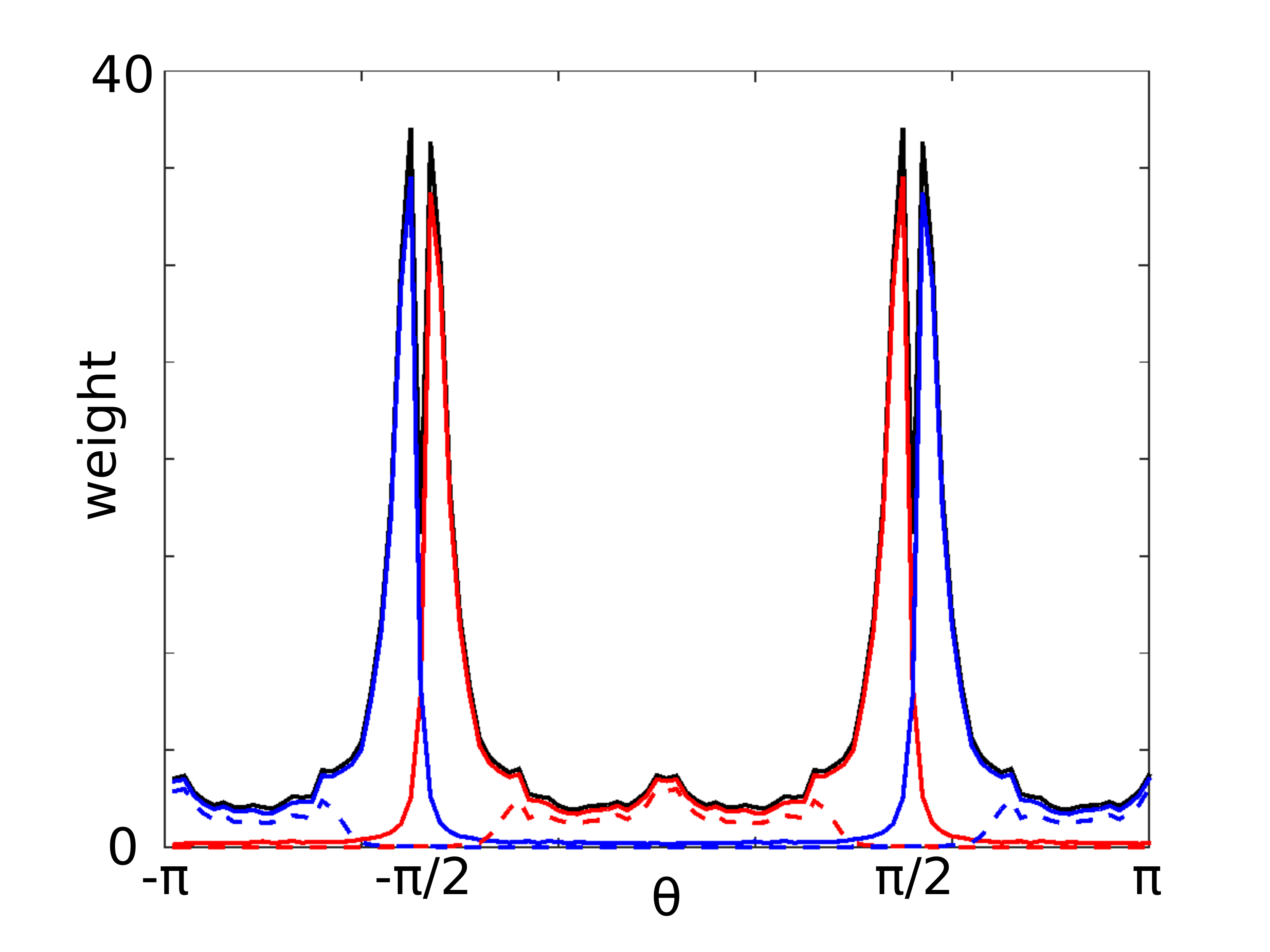}
\caption{Histogram of the collection of all $\theta_{\nu}(\mathbf{i})$, weighted by $|\delta_{\nu}(\mathbf{i})|$ for a conventional $s$-wave superconductor with a generic band structure and no spin-orbit coupling. The three distributions drawn with full lines correspond to contributions from quasiparticle states with $E_{\nu} < 0$ (red), $E_{\nu} > 0$ (blue), and all states (black).
The dashed lines include only contributions from quasiparticle states close to the Fermi level, $|E_{\nu}| < 2|\Delta|$.}
\label{Figure:PhaseHistogram}
\end{figure}

In Fig.~\ref{Figure:PhaseHistogram} we plot a histogram of $\theta_{\nu}(\mathbf{i})$, taken over all $\nu$ and $\mathbf{i}$ and weighted by $|\delta_{\nu}(\mathbf{i})|$, for a conventional $s$-wave superconductor.
As seen, the quasiparticle states with $E_{\nu} < 0$ (red lines) and $E_{\nu} > 0$ (blue lines) to a very high degree satisfies $|\theta_{\nu}(\mathbf{i})| < \pi/2$ and $|\theta_{\nu}(\mathbf{i})| > \pi/2$, respectively.
That is, states below the Fermi level tend to be overall more in-phase with the condensate at each point in space, while those above tend to be more out-of-phase with it.
In fact, states close to the Fermi level (dashed lines) have phases much closer to $\theta = 0$ for occupied states and $\theta = \pm\pi$ for unoccupied states, than those that are further away from the Fermi level.

\begin{figure}
\includegraphics[width=245pt]{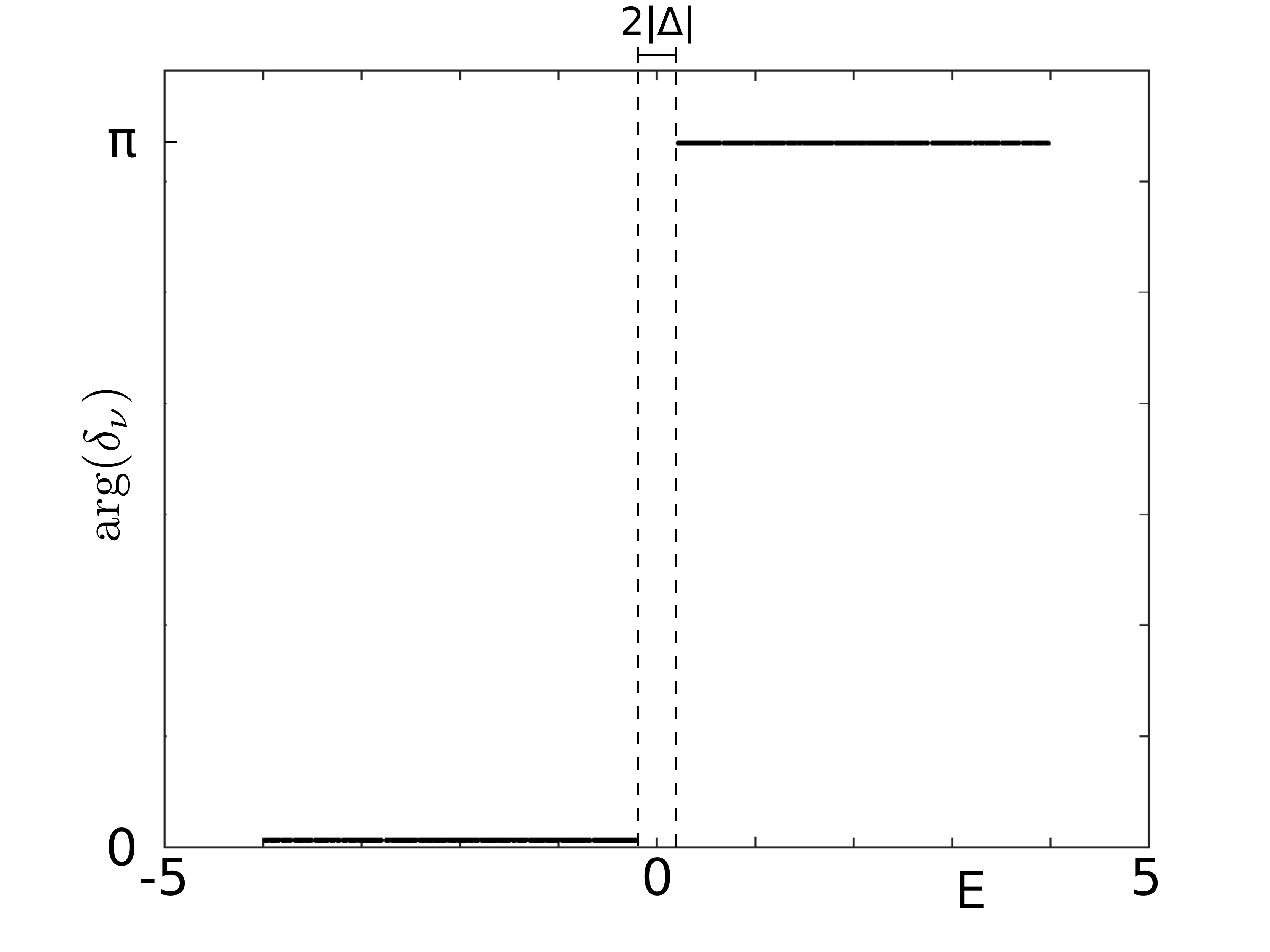}
\caption{The point ($E_{\nu}$, $\textrm{arg}(\delta_{\nu}))$) plotted for each eigenstate of a conventional $s$-wave superconductor with a generic band structure and no spin-orbit coupling.}
\label{Figure:PhaseAverage2}
\end{figure}

\begin{figure}
\includegraphics[width=145pt]{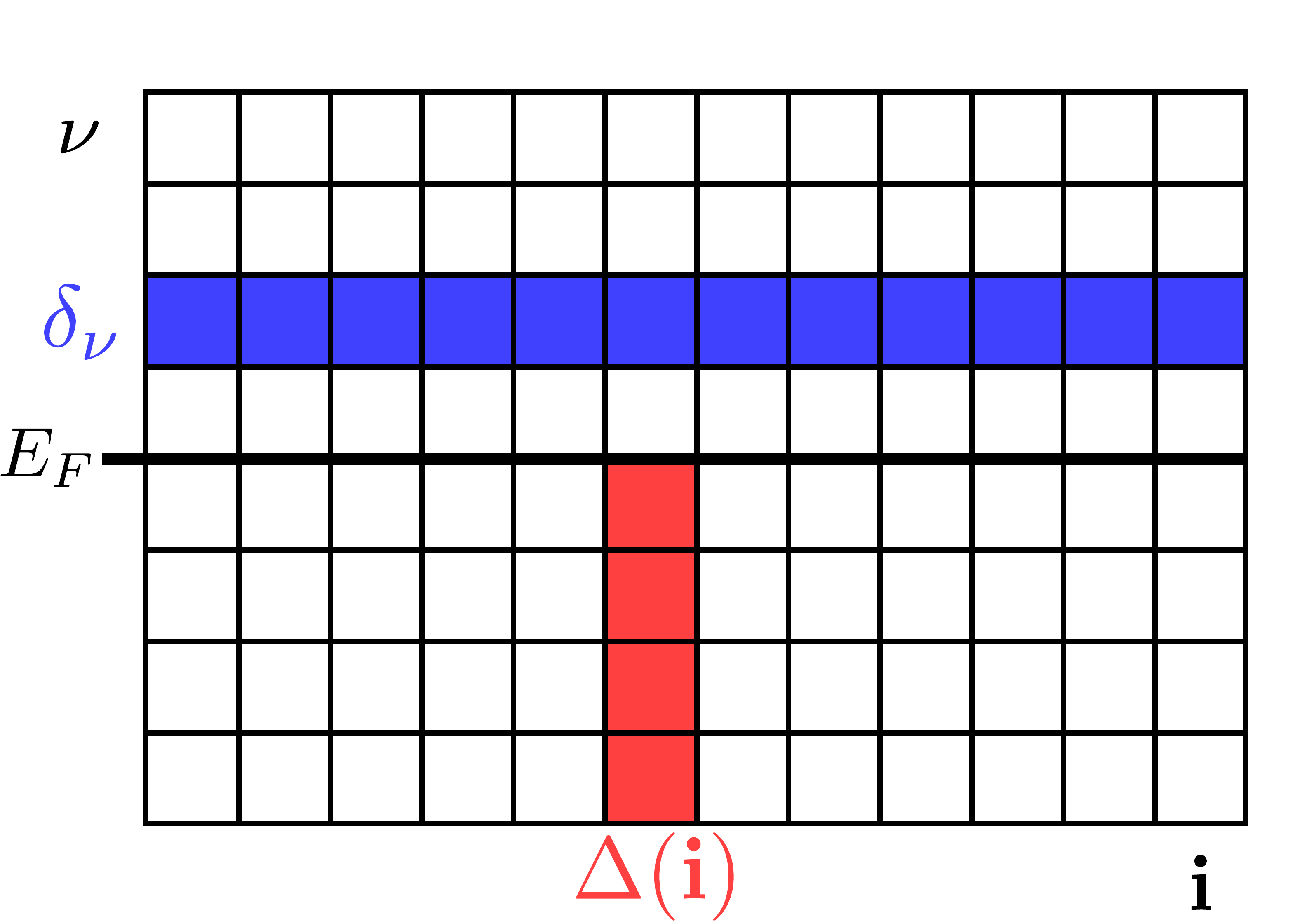}
\caption{Schematic view of the relation between $\delta_{\nu}(\mathbf{i}), \delta_{\nu}$, and $\Delta(\mathbf{i})$.
Rows correspond to eigenstates, while columns corresponds to sites in space, with each square corresponding to one $\delta_{\nu}(\mathbf{i})$.
The parameter $\delta_{\nu}$ that is used to classify eigenstates as being in- or out-of-phase with the condensate is obtained by summing horizontally over all spatial indices in a single eigenstate, while the order parameter $\Delta(\mathbf{i})$ is obtained by summing vertically over all eigenstate indices below the Fermi level.
}
\label{Figure:Schematic}
\end{figure}

We can also sum the distributions in Fig.~\ref{Figure:PhaseHistogram} over either eigenstate or site index.
If we sum over all occupied eigenstates $\nu$, the non-zero phases average out and give rise to a non-zero order parameter with phase $\theta = 0$ at each site $\mathbf{i}$, as we expect for a conventional $s$-wave superconductor.
This is despite the fact that each distribution peaks around $\pm \pi/2$.
Interestingly, if we instead sum over all sites $\mathbf{i}$:
\begin{align}
\label{Equation:Average_delta}
	\delta_{\nu} =& \frac{1}{N}\sum_{\mathbf{i}}\delta_{\nu}(\mathbf{i}),
\end{align}
where $N$ is the number of lattice sites, the final phase is either 0 or $\pi$, depending on the state being below or above the Fermi level, respectively, as we show in Fig.~\ref{Figure:PhaseAverage2}.
That is, even though each occupied quasiparticle state is not necessarily in-phase with the condensate at every point in space, it is still perfectly in-phase if spatially averaged.
Based on this observation, we can use the quantity $\textrm{sgn}\left[\textrm{Re}\left(\delta_{\nu}\right)\right]$ to classify individual quasiparticle states as on average (in space) being in- or out-of-phase with the condensate.
A schematic diagram over the relation between $\delta_{\nu}(\mathbf{i})$, $\delta_{\nu}$, and $\Delta(\mathbf{i})$ can be seen in Fig.~\ref{Figure:Schematic}.

Further motivation for, and insight into, why it is the real part of $\delta_{\nu}$ that is of interest can be obtained by considering the energy of a single quasiparticle state, which can be written as
\begin{align}
\label{Equation:Ev}
	E_{\nu} = &\int d\mathbf{x}\left[\begin{array}{cc}
			\mathbf{u}_{\nu\mathbf{x}}^{\dagger}	& \mathbf{v}_{\nu\mathbf{x}}^{\dagger}
		\end{array}\right]\mathcal{H}_{BdG}(\mathbf{x})\left[\begin{array}{c}
			\mathbf{u}_{\nu\mathbf{x}}\nonumber\\
			\mathbf{v}_{\nu\mathbf{x}}
		\end{array}\right]\\
		=& \int d\mathbf{x}\left(\mathbf{u}_{\nu\mathbf{x}}^{\dagger}\mathbf{H}_{0}(\mathbf{x})\mathbf{u}_{\nu\mathbf{x}} - \mathbf{v}_{\nu\mathbf{x}}^{\dagger}\mathbf{H}_{0}^{T}(\mathbf{x})\mathbf{v}_{\nu\mathbf{x}}\right)\nonumber\\
		&+ \int d\mathbf{x}\left(\mathbf{u}_{\nu\mathbf{x}}^{\dagger}\boldsymbol{\Delta}(\mathbf{x})
			\mathbf{v}_{\nu\mathbf{x}} + \mathbf{v}_{\nu\mathbf{x}}^{\dagger}\boldsymbol{\Delta}^{\dagger}(\mathbf{x})\mathbf{u}_{\nu\mathbf{x}}\right).
\end{align}
It is clear that the first term in Eq.~\eqref{Equation:Ev} is the energy contribution from fractionally occupied electron and hole states.
The second term can be rewritten as
\begin{align}
\label{Equation:Resonance_energy}
	&\int d\mathbf{x}\Delta(\mathbf{x})\left(u_{\nu\mathbf{x}\uparrow}^{*}v_{\nu\mathbf{x}\downarrow} - u_{\nu\mathbf{x}\downarrow}^{*}v_{\nu\mathbf{x}\uparrow}\right)\nonumber\\
	&+ \int d\mathbf{x}\Delta^{*}(\mathbf{x})\left(v_{\nu\mathbf{x}\downarrow}^{*}u_{\nu\mathbf{x}\uparrow} - v_{\nu\mathbf{x}\uparrow}^{*}u_{\nu\mathbf{x}\downarrow}\right)\nonumber\\
	=& 2\int d\mathbf{x}|\Delta(\mathbf{x})|^2\textrm{Re}\left(\frac{v_{\nu\mathbf{x}\downarrow}^{*}u_{\nu\mathbf{x}\uparrow} - v_{\nu\mathbf{x}\uparrow}^{*}u_{\nu\mathbf{x}\downarrow}}{\Delta(\mathbf{x})}\right).
\end{align}
From this it directly seen that the quasiparticle states get an additional energy contribution that is negative or positive depending on whether they are in- or out-of-phase with the condensate, respectively.
We can thus think of this as a resonance energy for the quasiparticle state.
With this insight it is clear that the expression in Eq.~\eqref{Equation:Average_delta} is very closely related to the total resonance energy of the quasiparticle state.
Note, however, that $\delta_{\nu}(\mathbf{i})$ has been defined with the bulk phase of the superconducting order parameter in the denominator in order to be able to determine whether the state is in- or out-of-phase with the whole bulk condensate.
In contrast, the term that enters into the resonance energy in Eq.~\eqref{Equation:Resonance_energy} has the local order parameter in the denominator.
In cases where the order parameter varies in space, it is really whether the state is in- or out-of-phase with the local order parameter that is of importance for the energetics.
However, here we are primarily interested in the phase relative to the surrounding bulk, in which case Eq.~\ref{Equation:Average_delta} is the relevant quantity.

\section{Ferromagnetic wire}
Previous studies of single magnetic impurities in conventional $s$-wave superconductors have found a sign change, or $\pi$-shift for the superconducting order parameter locally at the impurity site \cite{RevModPhys.78.373, PhysRevB.55.12648, PhysRevLett.78.3761, PhysRevB.92.064503}.
Here we extend this study to a one-dimensional wire of ferromagnetically aligned magnetic impurities. As it turns out, several impurities results in a richer behavior and is also more helpful for understanding the phenomenon itself. Ferromagnetic wires can also be used as the basic building block for engineering various types of $\pi$-junctions. Moreover, by also allowing for a finite spin-orbit coupling, which we do in the next section, a non-trivial one-dimensional topological state can be achieved, which host Majorana fermions at the wire end points, and is a system that have recently generated a significant amount of interest.\cite{PhysRevLett.105.077001, PhysRevLett.105.177002, Science.336.1003, NatPhys.8.795, NatPhys.8.887, Science.346.602, arXiv:1505.06078, PhysRevLett.115.197204, PhysRevB.88.020407, PhysRevB.84.195442}

\begin{figure}[htb]
\includegraphics[width=245pt]{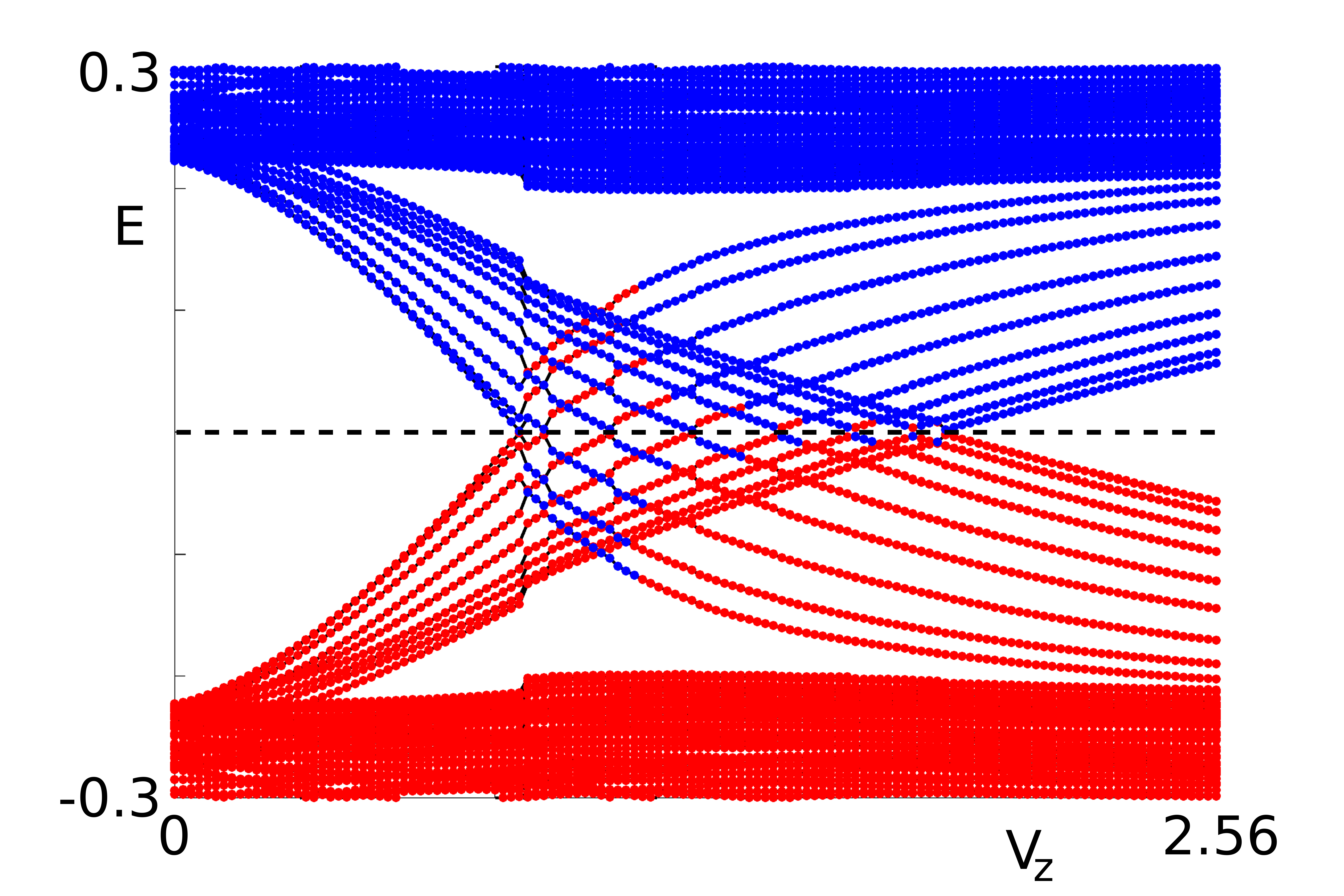}
\caption{Energy spectrum as a function of magnetic impurity strength $V_z$ for a ferromagnetic wire embedded in a two-dimensional conventional $s$-wave superconductor without spin-orbit coupling.
States in-phase with the condensate are marked in red i.e.~when $\textrm{sgn}\left[\textrm{Re}\left(\delta_{\nu}\right)\right] > 0$, while those out-of-phase with the condensate are marked in blue, i.e.~when $\textrm{sgn}\left[\textrm{Re}\left(\delta_{\nu}\right)\right] < 0$.}
\label{Figure:Wire_spectrum}
\end{figure}

\begin{figure}[htb]
\includegraphics[width=245pt]{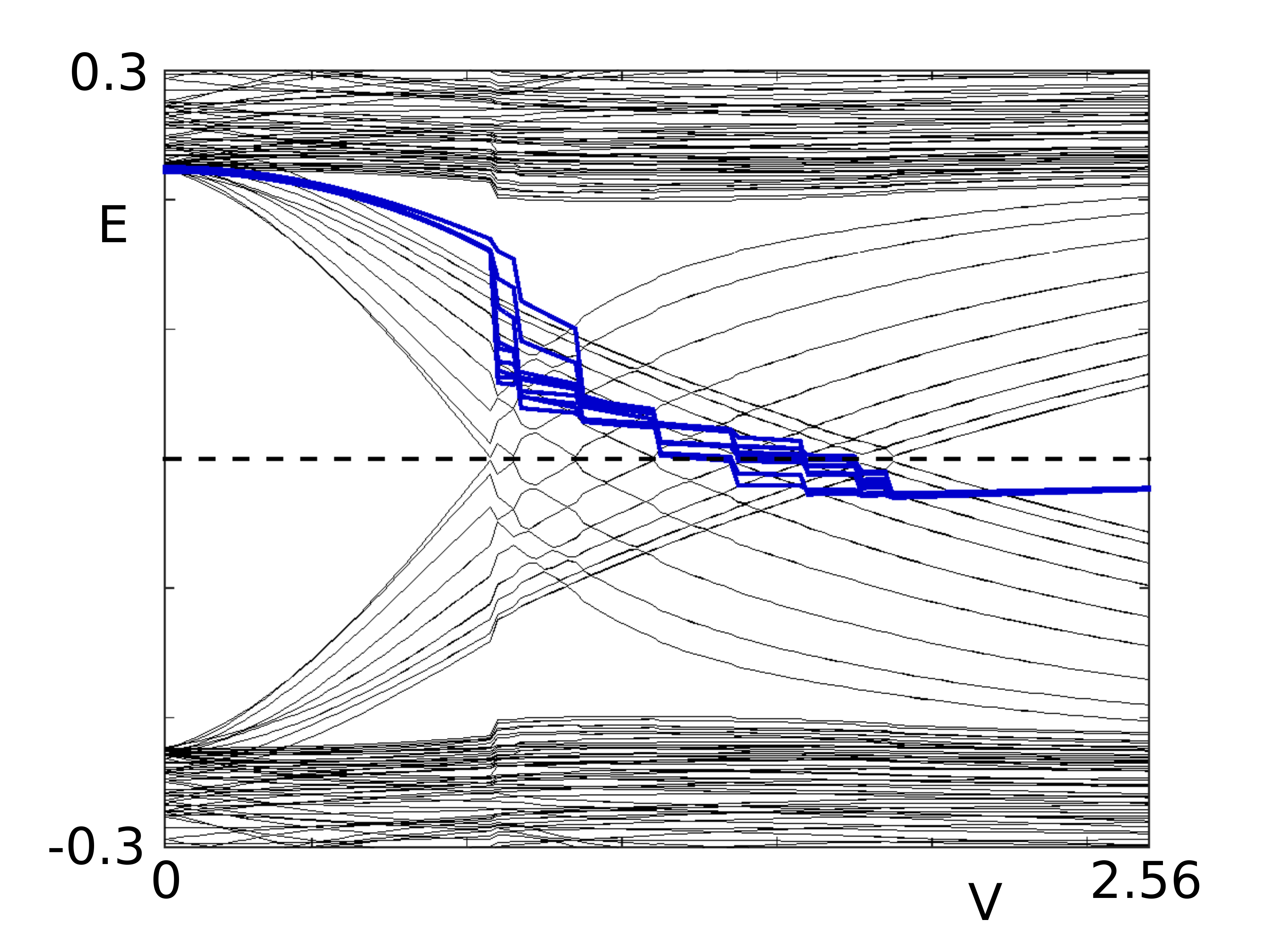}
\caption{Same energy spectrum as a function of magnetic impurity strength $V_z$ as in Fig.~\ref{Figure:Wire_spectrum} (thin black lines) and the real part of the superconducting order parameter at each site of the wire (thick blue lines).
Note that we allow for an overall complex order parameter and the blue lines are therefore the real part of the order parameters after having been divided by the phase $\Delta/|\Delta|$ of the bulk condensate, similarly as for $\delta_{\nu}(\mathbf{i})$.}
\label{Figure:D}
\end{figure}

\begin{figure}[htb]
\includegraphics[width=245pt]{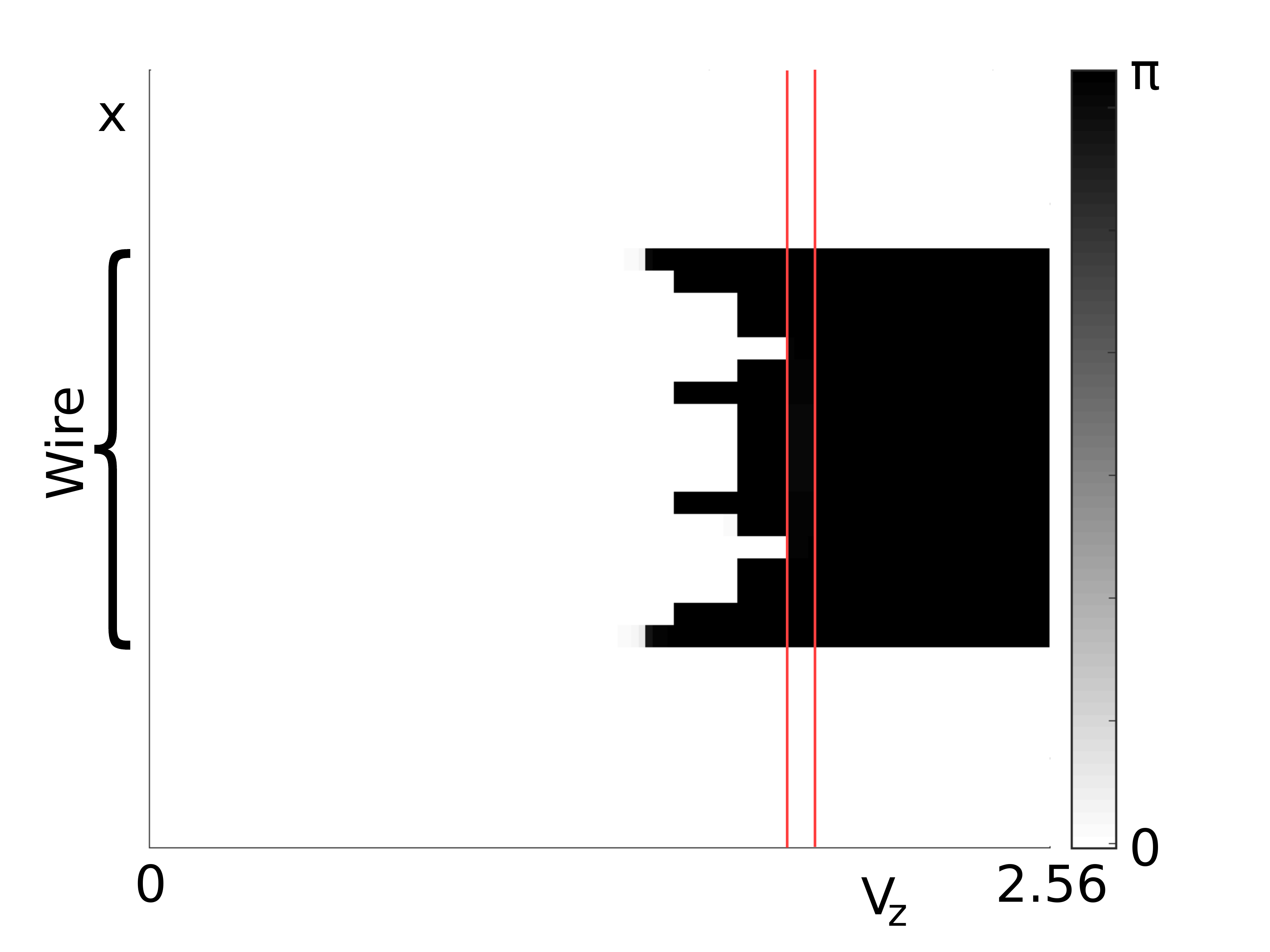}
\caption{Phase of the superconducting order parameter on the sites along a straight line going through the magnetic impurity wire, as a function of the magnetic impurity strength $V_z$.
The region between the two red vertical lines is discussed further in Sec.~\ref{Section:Phase_gradient_and_currents}.}
\label{Figure:D2}
\end{figure}

In Fig.~\ref{Figure:Wire_spectrum} we plot the energy spectrum for a ferromagnetic impurity wire embedded in a two-dimensional conventional $s$-wave superconductor and mark the eigenstates red or blue according to whether they are in- or out-of-phase with the condensate, respectively, using the quantity $\textrm{sgn}\left[\textrm{Re}\left(\delta_{\nu}\right)\right]$.
It is clear that for $V_z = 0$, all states below the Fermi level are in resonance with the condensate, in natural agreement with the results in Fig.~\ref{Figure:PhaseAverage2}.
As the strength of the Zeeman term on the wire is increased, the YSR impurity states are pulled deeper into the gap and towards the Fermi level.
Eventually, these states start to cross the Fermi level. Each time this happens a state that is in-phase with the condensate is deoccupied, while another state that is out-of-phase with the condensate becomes occupied.
It is therefore natural to expect that the order parameter suddenly decreases at each such Fermi level crossing. In Fig.~\ref{Figure:D} we plot the local superconducting order parameter at each site of the wire (blue lines), overlayed on the energy spectrum (thin black lines), and it is clear that the local order parameters indeed exhibit discrete downwards jumps at each Fermi level crossing.
Each time a YSR state crosses the Fermi level, the order parameter makes a jump, but the magnitude of the jump is different depending on wire site.
This can easily be understood since the wave function of the quasiparticle states have different amplitudes on different sites, and thus its influence on the order parameter is naturally site dependent.
At strong enough Zeeman fields the order parameter on the wire changes sign compared to the surrounding bulk condensate and thus there is a local $\pi$-shift concentrated along the ferromagnetic impurity wire.
As can be expected from the different sizes of the jumps in the order parameter, this does not happen at each wire site for one fixed Zeeman term, but rather different sites acquires a $\pi$-shifted order parameter at slightly different magnetic strengths. We illustrate this in Fig.~\ref{Figure:D2} where we plot the phase of the superconducting order parameter on the sites along a line going through the magnetic wire as function of the magnetic strength $V_z$.
Clearly, the superconducting order parameter is $\pi$-phase shifted for the whole wire for large enough $V_z$, but there is a finite transition region where some, but not all, sites along the wire experience a $\pi$-shift. We also note that the $\pi$-shift is highly localized to the wire sites, and does not spread even to the neighboring sites. The $\pi$-shift for a single impurity has previously been found to be of similarly short-ranged character, with an extent that is proportional to the Fermi wave length $k_F$.\cite{PhysRevLett.78.3761}

We finally also make a note about the change of phase that occurs for the YSR states beyond the Fermi level crossing at high $V_z$, as seen in Fig.~\ref{Figure:Wire_spectrum}.
It is clear that the real part of $\delta_{\nu}$ eventually changes sign to that of the condensate again, as the states are pulled deep enough below the Fermi level.
Naively this change of sign could be expected to give rise to a compensatory upward jump in the order parameter.
However, we remember that $\delta_{\nu}$ is calculated as an average of $\delta_{\nu}(\mathbf{x})$ over the whole sample.
The change in sign here merely reflects that the state in some spatial regions becomes in-phase with the bulk condensate but it can still stay out-of-phase in other regions, which is exactly what happens for the wire sites. 

\section{Rashba spin-orbit interaction}
\begin{figure}[htb]
\includegraphics[width=245pt]{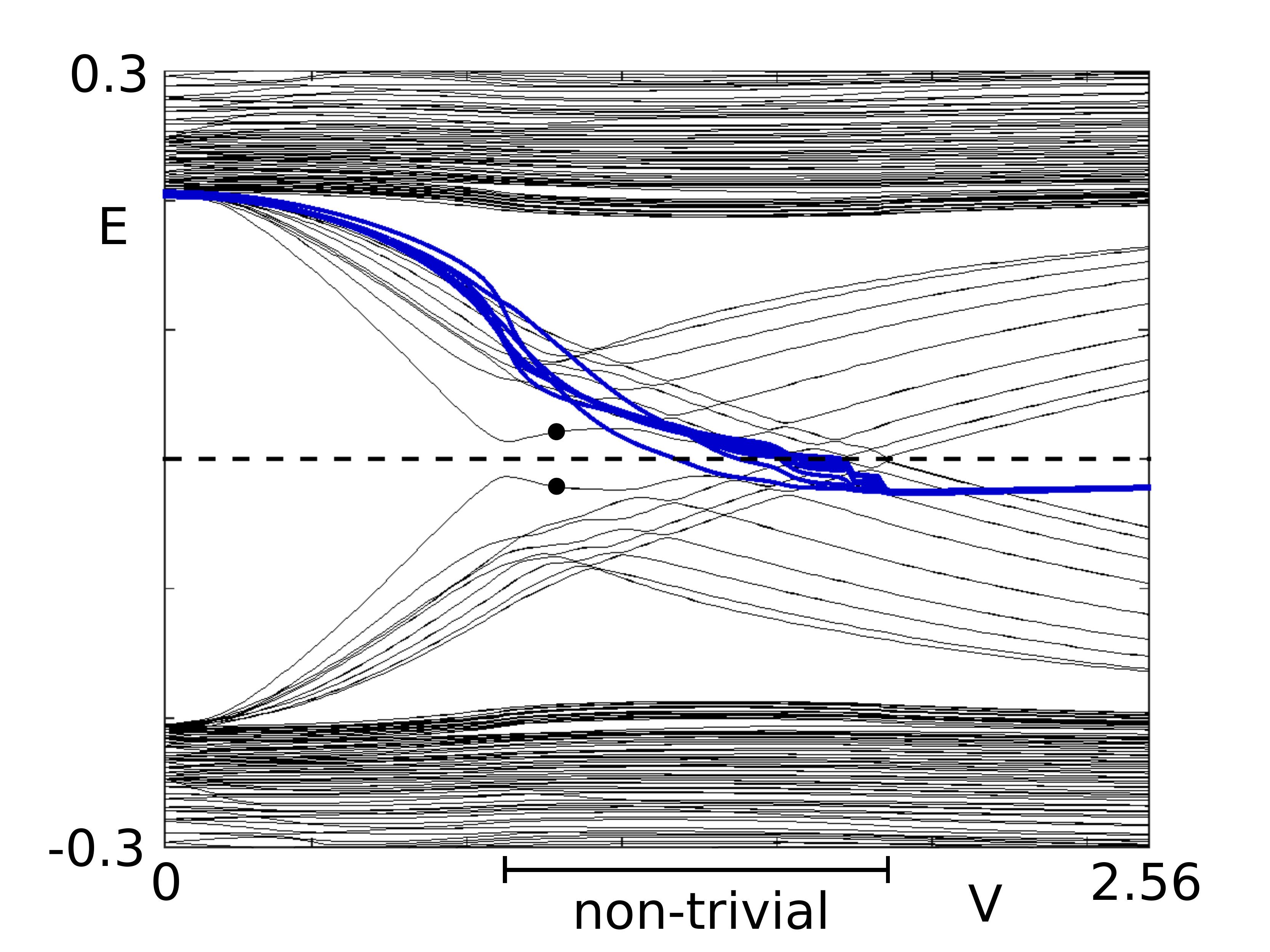}
\caption{Same as in Fig.~\ref{Figure:D}, but for finite spin-orbit coupling $\alpha = 0.3$.
The two dots on the lines closest to zero corresponds to eigenvalues for which 
the combined probability density is plotted in Fig.~\ref{Figure:LowestEnergyDensity}.}
\label{Figure:D_SO}
\end{figure}

\begin{figure}
\includegraphics[width=245pt]{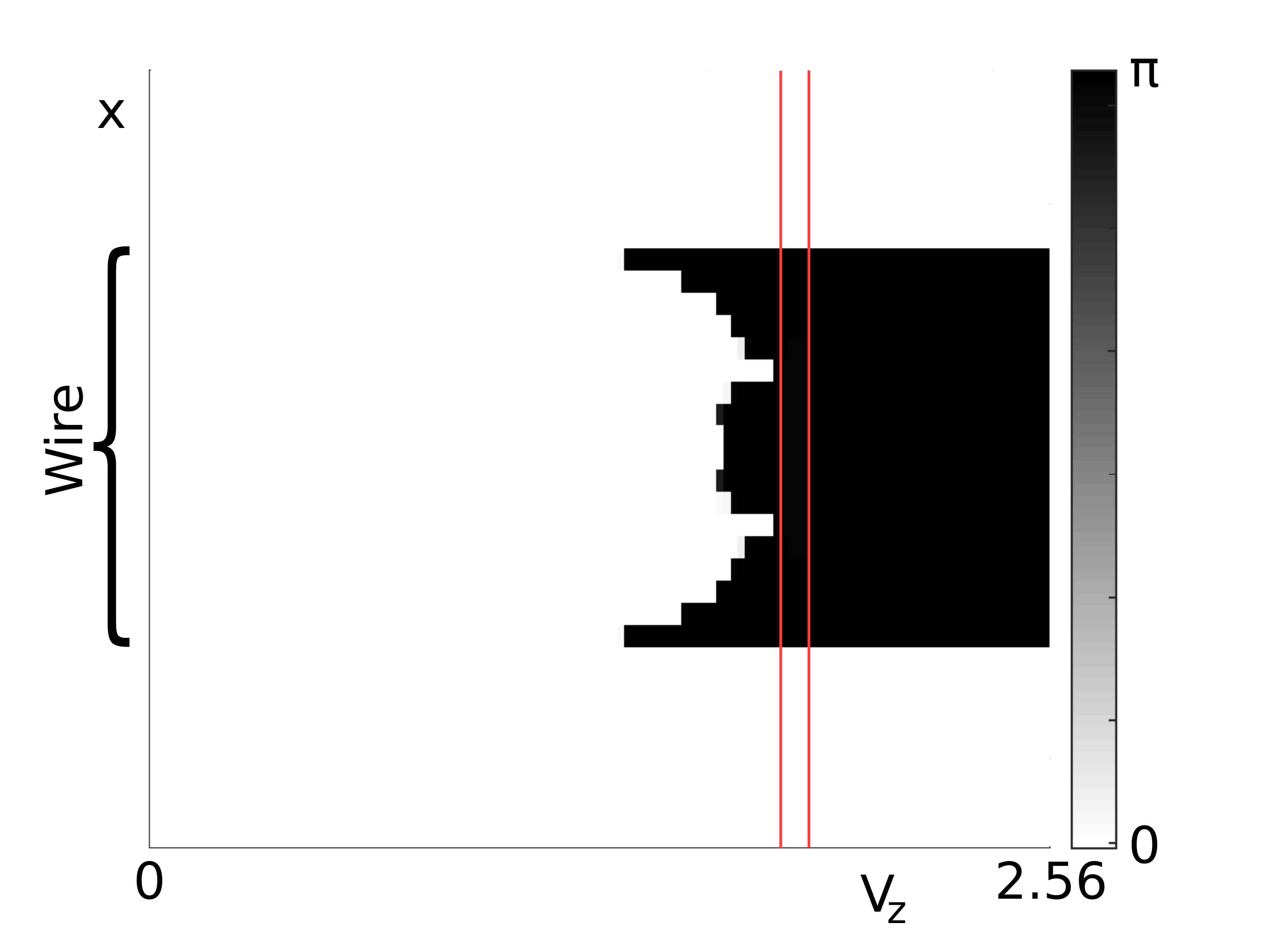}
\caption{Same as in Fig.~\ref{Figure:D2}, but for finite spin-orbit coupling $\alpha = 0.3$.}
\label{Figure:D2_SO}
\end{figure}

We now turn to the situation where we allow for a finite Rashba spin-orbit interaction in the superconductor. For a range of $V_z$ values the wire now enters a non-trivial one-dimensional superconducting topological phase.\cite{PhysRevLett.105.077001, PhysRevLett.105.177002, Science.336.1003, NatPhys.8.795, NatPhys.8.887, Science.346.602, arXiv:1505.06078, PhysRevLett.115.197204, PhysRevB.88.020407, PhysRevB.84.195442}
The YSR states moving through the energy gap from above and below do not couple to each other when only $s$-wave superconductivity and a Zeeman term is present.
However, once a Rashba spin-orbit interaction is introduced, the states can couple to each other, which tends to push these intragap states away from the Fermi level.\cite{arXiv:1605.00696}
As is clearly visible in Fig.~\ref{Figure:D_SO}, it also leads to avoided crossings.
Moreover, the character of the YSR states above and below the Fermi level are now mixed with each other as the strength of the Zeeman term increases. Thus, rather than the order parameter experiencing sudden discrete jumps as the states changes their occupation, it now changes continuously with the changing character of the occupied quasiparticle states.
This leads to a qualitatively similar decrease in the superconducting order parameter, and the eventual $\pi$-shift with increasing $V_z$, as for no spin-orbit interaction, except for the absence of the discrete jumps. 
The qualitatively similar behavior of the order parameter along the wire is also direct visible in a comparison between Fig.~\ref{Figure:D2}, which is without spin-orbit coupling, and Fig.~\ref{Figure:D2_SO} which is for a finite spin-orbit coupling.

\begin{figure}
\includegraphics[width=245pt]{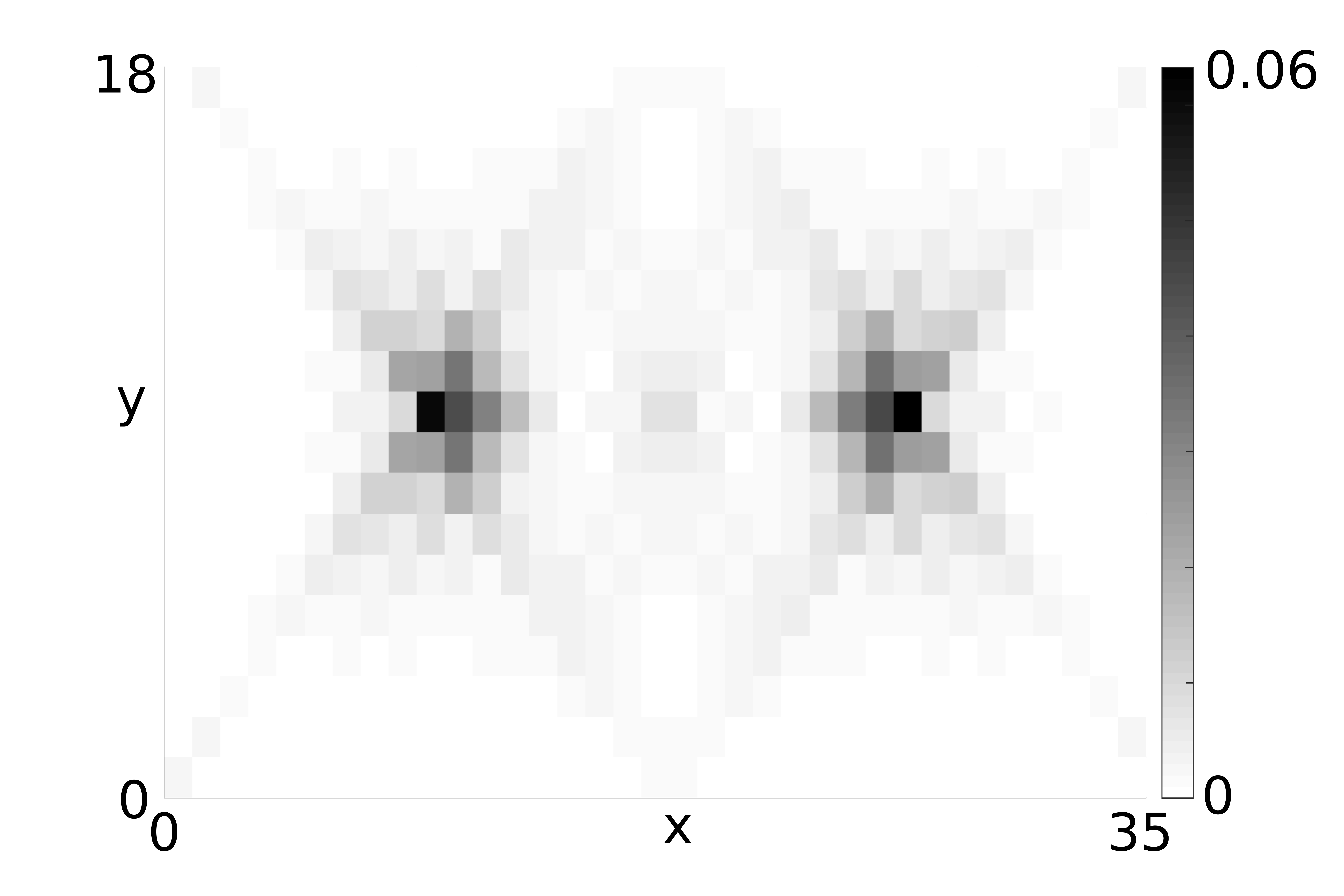}
\caption{The combined probability density for the two states closest to $E = 0$ at $V_z = 1$, as indicated by dots in Fig.~\ref{Figure:D_SO}, clearly localized around the two wire end points, as expected for two Majorana fermions.}
\label{Figure:LowestEnergyDensity}
\end{figure}

Although we in this work are not primarily concerned with topological superconductivity, we nonetheless marked the non-trivial region in Fig.~\ref{Figure:D_SO}.
The two states closest to the Fermi level in this regime corresponds to the single Majorana fermions at the two wire end points, which are here somewhat hybridized due to a short wire length. This can be verified by looking at the probability density of these two states, which is clearly peaked at the wire end points as displayed in Fig.~\ref{Figure:LowestEnergyDensity}. We also note that the $\pi$-shift of the superconducting order parameter takes place at significantly higher values of $V_z$ than the topological phase transition, and are therefore not directly related to each other.

\section{Phase gradient and currents}
\label{Section:Phase_gradient_and_currents}
So far we have focused on the origin and occurrence of the $\pi$-shift of the order parameter at the ferromagnetic impurity sites. We now finish with an investigation of any possible phase gradients in the superconducting order parameter in the regions surrounding the magnetic wire. This is particularly relevant as previous work has found persistent currents around ferromagnetic impurities in conventional $s$-wave superconductors in the presence of spin-orbit interaction.\cite{PhysRevLett.115.116602, PhysRevB.92.214501}
We therefore carefully examine the phase of the superconducting order parameter in the whole sample, paying particular attention to not bias the self-consistency procedure to a real solution. Even when starting the self-consistency loop with a complex phase that is randomly varied from site to site over the phase range of $\pi/10$, we find {\it no} phase gradients in the whole sample (apart from the discrete $\pi$-shift on the wire), for almost all values of the Zeeman exchange field $V_z$ of the impurities. It is only in a very narrow region in $V_z$-space, just after the whole wire has transitioned into the $\pi$-shifted state, that we find some minor phase gradients in the order parameter. The boundaries of the region with finite phase gradients are indicated by vertical red lines in Figs.~\ref{Figure:D2} and \ref{Figure:D2_SO}, and are largely independent on the spin-orbit coupling. 

\begin{figure}.
\includegraphics[width=245pt]{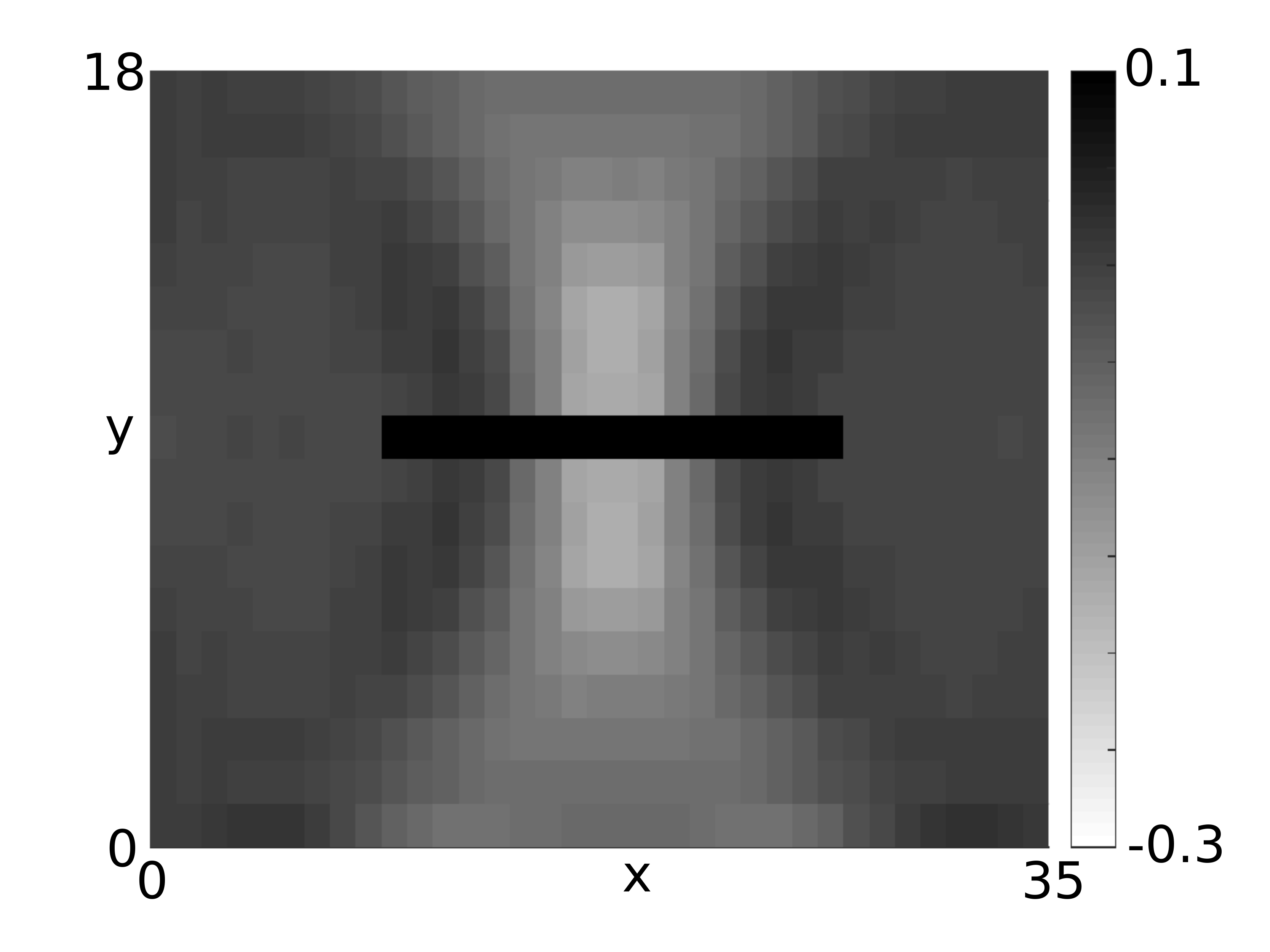}
\caption{Phase of the order parameter over the whole system, for $V_z = 1.82$ and $\alpha = 0$.
The plot is representative of the phase behavior for all values inside the regions delimited by red vertical lines in both Figs.~\ref{Figure:D2} and \ref{Figure:D2_SO}.
Note that the phase variations are very small and that the values on the wire itself is $\pi$ rather than $0.1$ as the color scale has been limited to $[-0.3, 0.1]$.}
\label{Figure:D_complete}
\end{figure}

In Fig.~\ref{Figure:D_complete} we plot a representative view of the superconducting phase inside this region for zero spin-orbit coupling, but the picture is, even quantitatively, the same for finite spin-orbit coupling.
As seen, although there is a finite gradient, the difference between the maximum and minimum phase in the region surrounding the wire is less than $0.4$ and is therefore small compared to for example those surrounding a vortex, where the phase of the order parameter twists by a full $2\pi$. The resulting currents carried by the superconducting condensate is thus necessarily quite small.
Moreover, the phase gradient sets the direction of motion of the Cooper pairs, with the bright and dark spots in the region surrounding the wire thus corresponding to sinks and sources of the current carried by the condensate.
While the total current is continuous, as shown in Ref.~[\onlinecite{PhysRevB.92.214501}], this reflects the fact that only part of the total current is actually carried by the superconducting condensate.
It is true that a phase gradient in the superconducting order parameter gives rise to a current, related to the center of mass motion of the Cooper pairs, but this is only one possible part of the total current, as also single quasiparticles can carry current. 
Thus, the persistent currents found for magnetic impurities in spin-orbit coupled superconductors can according to our results not be related to the superconducting order parameter developing phase gradients; we only find finite phase gradients in a narrow window of $V_z$ and independent on the presence of spin-orbit coupling, while persistent currents are present for all finite $V_z$ but only for finite spin-orbit coupling. We instead interpret the small phase gradients as a consequence of a system instability around the $\pi$-shift region and independent on the system hosting persistent currents.

\section{Discussion and conclusions}
In this work we have studied a wire built of ferromagnetic impurities embedded in a conventional $s$-wave superconductor.  
With increasing magnetic impurity strength, the energy of the impurity-induced YSR states decreases and eventually crosses the Fermi level. We find that these zero-energy crossings result in discrete downward jumps of the superconducting order parameter on the impurity sites, eventually leading to a $\pi$-phase shift of the superconducting order parameter in the whole wire. 
By establishing a way to evaluate how well individual quasiparticle states are in-phase, or in resonance, with the overall superconducting condensate, we can understand these jumps as a direct consequence of the non-resonating character for initially unoccupied YSR states, which once they are pulled below the Fermi level contribute destructively to the order parameter.
However, it is clear that pulling a single YSR state below the Fermi level does not necessarily result in a $\pi$-phase shift of the order parameter. Rather, such a $\pi$-shift only occurs once the out-of-phase states dominate the local contribution to the order parameter. This can easily happen at different magnetic impurity strengths for neighboring sites in the same magnetic wire.
We also find that the behavior of the superconducting order parameter is qualitatively the same even in the presence of finite Rashba spin-orbit interaction, except that the discrete jumps are now smooth, as the quasiparticle states mix around the Fermi level.
In addition, we establish that the superconducting order parameter only hosts finite phase gradients in and around the ferromagnetic wire in a very narrow window around the $\pi$-shift. These phase gradients are small, and are not related to the persistent currents previously found for magnetic impurities in spin-orbit coupled superconductors.\cite{PhysRevLett.115.116602, PhysRevB.92.214501}
Finally, we point out that Figs.~\ref{Figure:D2} and Fig.~\ref{Figure:D2_SO} clearly show that it is possible to engineer various types of $\pi$-junctions by tuning the strength of the magnetic impurities.

\section{Acknowledgement}
We thank B.~M.~Andersen, B.~A.~Bernevig, and J.~Paaske for interesting discussions.
This work was supported by the Swedish Research Council (Vetenskapsr\aa det), the Swedish Foundation for Strategic Research (SSF), the G\"{o}ran Gustafsson Foundation, the Wallenberg Academy Fellow program, and the European Research Council (ERC DM-321031).
The computations were performed on resources provided by SNIC through Uppsala Multidisciplinary Center for Advanced Computational Science (UPPMAX) under Project snic2016-1-19.
Work at Los Alamos was supported by the US DoE BES E304.


\begin{thebibliography}{38}
	\bibitem{JPhysChemSolids.11.26}
		P. W. Anderson,
		J. Phys. Chem. Solids {\bf 11}, 26 (1959).
	\bibitem{RevModPhys.78.373}
		A. V. Balatsky, I. Vekhter, and J.-X. Zhu,
		Rev. Mod. Phys. {\bf 78}, 373 (2006).
	\bibitem{ActaPhysSin.21.75}
		L. Yu,
		Acta Phys. Sin. {\bf 21}, 75 (1965).
	\bibitem{ProgTheorPhys.40.435}
		H. Shiba,
		Prog. Theor. Phys. {\bf 40}, 435 (1968).
	\bibitem{JETPLett.9.85}
		A. I. Rusinov,
		JETP Lett. {\bf 9}, 85 (1969).
	\bibitem{PhysRevB.55.12648}
		M. I. Salkola, A. V. Balatsky, and J. R. Schrieffer,
		Phys. Rev. B {\bf 55}, 12648 (1997).
	\bibitem{PhysRevLett.78.3761}
		M. E. Flatt\'e and J. M. Byers,
		Phys. Rev. Lett. {\bf 78}, 3761 (1997).
	\bibitem{PhysRevB.92.064503}
		T. Meng, J. Klinovaja, S. Hoffman, P. Simon, and D. Loss,
		Phys. Rev. B {\bf 92}, 064503 (2015).
	\bibitem{PhysRevB.88.155420}
		F. Pientka, L. I. Glazman, and F. von Oppen,
		Phys. Rev. B {\bf 88}, 155420 (2013).
	\bibitem{PhysRevLett.114.236803}
		J. R\"{o}ntynen and T. Ojanen,
		Phys. Rev. Lett. {\bf 114}, 236803 (2015).
	\bibitem{arXiv:1605.00696}
		K. Bj\"{o}rnson and A. M. Black-Schaffer,
		Phys. Rev. B {\bf 94}, 100501 (2016).
	\bibitem{PhysRevLett.105.077001}
		R. M. Lutchyn, J. D. Sau, and S. Das Sarma,
		Phys. Rev. Lett. {\bf 105}, 077001 (2010).
	\bibitem{PhysRevLett.105.177002}
		Y. Oreg, G. Refael, and F. von Oppen,
		Phys. Rev. Lett. {\bf 105}, 177002 (2010).
	\bibitem{PhysRevB.84.195442}
		T.-P. Choy, J. M. Edge, A. R. Akhmerov, and C. W. J. Beenakker,
		Phys. Rev. B {\bf 84}, 195442 (2011).
	\bibitem{Science.336.1003}
		V. Mourik, K. Zuo, S. M. Frolov, S. R. Plissard, E. P. A. M. Bakkers, and L. P. Kouwenhoven,
		Science {\bf 336}, 1003 (2012).
	\bibitem{NatPhys.8.795}
		L. P. Rokhinson, X. Liu, and J. K. Furdyna,
		Nat. Phys. {\bf 8}, 795 (2012).
	\bibitem{NatPhys.8.887}
		A. Das, Y. Ronen, Y. Most, Y. Oreg, M. Heiblum, and H. Shtrikman,
		Nat. Phys. {\bf 8}, 887 (2012).
	\bibitem{PhysRevB.88.020407}
		S. Nadj-Perge, I. K. Drozdov, B. A. Bernevig, and A. Yazdani,
		Phys. Rev. B {\bf 88}, 020407(R) (2013).
	\bibitem{Science.346.602}
		S. Nadj-Perge, I. K. Drozdov, J. Li, H. Chen, S. Jeon, J. Seo, A. H. MacDonald, B. A. Bernevig, and A. Yazdani,
		Science {\bf 346}, 602 (2014).
	\bibitem{arXiv:1505.06078}
		R. Pawlak, M. Kisiel, J. Klinovaja, T. Meier, S. Kawari, T. Glatzel, D. Loss, and E. Meyer,
		arXiv:1505.06078 (2015).
	\bibitem{PhysRevLett.115.197204}
		M. Ruby, F. Pientka, Y. Peng, F. von Oppen, B. W. Heinrich, and K. J. Franke,
		Phys. Rev. Lett. {\bf 115}, 197204 (2015).
	\bibitem{PhysRevLett.115.116602}
		S. S. Pershoguba, K. Bj\"{o}rnson, A. M. Black-Schaffer, and A. V. Balatsky,
		Phys. Rev. Lett. {\bf 115}, 116602 (2015).
	\bibitem{PhysRevB.92.214501}
		K. Bj\"{o}rnson, S. S. Pershoguba, A. V. Balatsky, and A. M. Black-Schaffer,
		Phys. Rev. B {\bf 92}, 214501 (2015).		
	\bibitem{PhysRevLett.103.020401}
		M. Sato, Y. Takahashi, and S. Fujimoto,
		Phys. Rev. Lett. {\bf 103}, 020401 (2009).
	\bibitem{PhysRevB.82.134521}
		M. Sato, Y. Takahashi, and S. Fujimoto,
		Phys. Rev. B {\bf 82}, 134521 (2010).
	\bibitem{PhysRevLett.104.040502}
		J. D. Sau, R. M. Lutchyn, S. Tewari, and S. Das Sarma,
		Phys. Rev. Lett. {\bf 104}, 040502 (2010).
	\bibitem{PhysRevB.84.180509}
		A. M. Black-Schaffer and J. Linder,
		Phys. Rev. B {\bf 84}, 180509(R) (2011).
	\bibitem{PhysRevB.88.024501}
		K. Bj\"{o}rnson and A. M. Black-Schaffer,
		Phys. Rev. B {\bf 88}, 024501 (2013).
	\bibitem{PhysRevB.91.214514}
		K. Bj\"{o}rnson and A. M. Black-Schaffer,
		Phys. Rev. B {\bf 91}, 214514 (2015).
\end{thebibliography}
\end{document}